\documentclass[showpacs,preprint,aps]{revtex4}

\usepackage{amsmath}
\usepackage{amsfonts}
\usepackage{amssymb}
\usepackage{graphicx}
\usepackage{color}
\usepackage{mathrsfs}

\newcommand{\be}{\begin{equation}}
\newcommand{\ee}{\end{equation}}

\newcommand{\W}{\Omega}

\newcommand{\ket}{\rangle}
\newcommand{\bra}{\langle}

\begin{document}

\title{Improvement of photovoltaic efficiency by Fano coherence}

\author{Dong Sun$^{1,2,4}$, Hui Long$^{1}$, Hongyi Lin$^{1,2}$, Yuri V. Rostovtsev$^{3,4}$}

\affiliation {$^{1}$ Department of Optoelectronics Engineering, Xiamen University of Technology, Xiamen 361024, China\\
$^{2}$ Fujian Provincial Key Laboratory of Optoelectronic technologies and devices, Xiamen 361024, China\\
$^{3}$ Department of Physics, University of North Texas, 1155 Union Circle \#311427, Denton, TX 76203\\
$^{4}$ Department of Physics, Texas A$\&$M University, College Station, TX 77843}

\date{\today}

\begin{abstract}
\vskip12 pt %%
We show that Fano resonance in the decay channels of a three-level system can lead to considerably absorption enhancement and emission suppression.
We found that a coherence built up in the ground doublet states, with strength depending on a coupling parameter that arises from the Fano
interference, can in principle lead to breaking of the detail balance between the absorption and emission processes in atomic systems.
\end{abstract}


\maketitle

\section{Introduction}

Quantum coherence effects, such as Coherent Population Trapping (CPT)~\cite{CPT} and Electromagnetically Induced Transparency
(EIT)\cite{book,harris97phys2day,harris93prl,fleisch05rmp}, have been the focus of broad research activities for the last two decades, as they
drastically change the optical properties of media. For example, for EIT in CW and pulsed regimes~\cite{harris97phys2day, harris93prl, fleisch05rmp,
ok86jetp,sau05pra}, absorption practically vanishes. Media with excited coherence may display high index of refraction without
absorption~\cite{scully91prl}. It is possible to achieve manipulation of a coherent medium~\cite{ye03optlett} and enhance nonlinear effects at a few
photon level~\cite{harris98prl,lukin00amop}.

The study of quantum interference and coherence effects in atomic and molecular systems \cite{kk1} has found numerous fascinating phenomena, e.g.,
Fano interference \cite{kk2}, vacuum induced coherence \cite{kk3}, lasing without inversion \cite{kk5,kk6}, quantum Carnot engine \cite{kk12} and
long lived coherences in biochemical molecules \cite{kk8}. The application of coherence in solar energy physics in fact can change the balance
processes that are limiting the operation of quantum systems \cite{kk13}. For example, for a quantum photocell, the fundamental limit to the
efficiency is accepted to be in the balance of radiative absorption and recombination. The coherence effects can in fact break this balance and
significantly suppress the emission process, resulting in enhancement of the power generated by photocell. One of the possible ways to break the
balance between recombination and absorption is via the coherent drive similar to the LWI process \cite{kk10,kk11}, where the coherence between two
levels induced by external source \cite{kk14} can cancel the emission processes. It is also possible to generate the coherence without using external
fields. This approach is based on the Fano effect \cite{kk2} that manifests itself as an interference between the eigenstates of the system. For
example, Fano interference was built up among states of two coupled quantum wells via tunneling \cite{kk16}. The direct application in an optical
system by means of lasing without inversion was analyzed in \cite{kk11} and has the name "Fano-Harris lasing without inversion" to distinguish it
from the standard (externally driven) LWI. The latest results showed that the quantum coherence that arises from the Fano coupling can significantly
enhance the power delivered to the load \cite{kk19}, as well as control over enhancement and suppression of the emission and absorption profiles
\cite{kk20}.

In this work we have studied the effect of Fano interference on the probabilities of emission and absorption. This interference is induced by two
spontaneous decays from discrete ground state doublet to an identical continuum.

The organization of this paper is as follows. In section II we discuss the theoretical model of the three-level system, the dynamical evolution of
the system, and the probability of absorption and emission. In section III we simulate the effect of the Fano interference between the decay channels
on the probability of absorption and emission. In section IV we show some analytical calculations of the Fano interference in both the probability
amplitude and the density matrix approaches. Finally in section V we summarize our results.

\section{Theoretical model}

\begin{figure}
\centering
\includegraphics[width=0.5\linewidth]{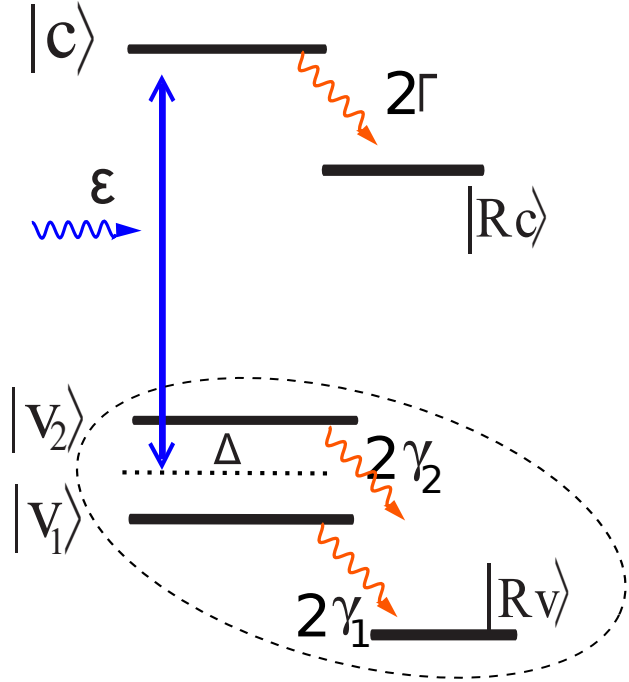}
\caption{Scheme of the three-level system with continuums. The Fano interference generates among the ground state doublet and reservoir state $R_v$ (enclosed by dashed ellipse).
$\cal{E}$ is the external weak electric field. $2\Gamma$ and $2\gamma_{1,2}$ are spontaneous decay rates from eigenstates to continuums.}\label{f1}
\end{figure}

An interesting example of Fano-like coupling is the three-level system as shown in Fig.\ref{f1}, where the effects of coherence play a major role.
This scheme is developed from the previous intersubband double quantum well structure \cite{kk31}. Consider a ground state doublet $|v_{1,2}\ket$ and
an excited state $|c\ket$ coherently driven by a weak electric field. This external field plays the role of the sunlight. We choose the central
frequency $\nu$ such that the energies of the state $|v_{1,2}\ket$ are related to $|c\ket$ as $\hbar(\nu \pm \Delta)$, where $\Delta$ is now the
frequency detuning, not the tunneling in \cite{kk31}. The ground state doublet decay to an identical continuum (we consider this continuum as a
reservoir state $R_v$) with rate $2\gamma_{1,2}$, and the excited state $|c\ket$ decays to a continuum (reservoir state $R_c$) with rate $2\Gamma$.

To get the dynamical evolution of this system, we can use both the probability amplitude method and the density matrix method. In the probability
amplitude method, we can write the state vector as
\begin{equation}
|\Psi\ket=v_1|v_1\ket+v_2|v_2\ket+c|c\ket.
\end{equation}
The dynamical equations of amplitude can be derived with the Weisskopf-Wigner approximation, which are also given in \cite{kk31} for intersubband double
quantum well structure by simply adding the decay term of $2\Gamma$ from the excited state $|c\ket$ to the reservoir $|R_c\ket$,
\begin{align}
\dot{v}_2 &=-(\gamma_2+i\Delta)v_2-p\sqrt{\gamma_1\gamma_2} v_1-i\Omega_2c \label{v2}, \\
\dot{v}_1 &=-(\gamma_1-i\Delta)v_1-p\sqrt{\gamma_1\gamma_2} v_2-i\Omega_1c \label{v1},\\
\dot{c} &=-i\Omega_2v_2-i\Omega_1v_1-\Gamma c \label{cc},
\end{align}
where $\Omega_{1,2}={\wp_{cv_{1,2}}\varepsilon}/{\hbar}$ are the Rabi frequencies of the applied field. $\wp_{cv_1}(\wp_{cv_2})$ is the dipole moment
of the transition $|c\ket \leftrightarrow |v_1\ket $ ($|c\ket \leftrightarrow |v_2\ket$), and $\mathcal{E}_0$ is the applied electric field. The
terms containing the product of the decay rates appear due to the interference introduced by the decay of the two optical transitions to the same
state. This is the so-called Fano interference, which couples the doublet states. In order to measure the strength of the interference, we introduce
the $p$ factor, which is the normalized scalar product of the corresponding dipole moments: \be
p=\frac{\wp_{R_vv_1}\cdot\wp_{R_vv_2}}{|\wp_{R_vv_1}\wp_{R_vv_2}|}.\ee

According to its definition, the alignment factor takes value $1$ for parallel dipole moments and $-1$ for antiparallel dipole moments, both
corresponding to the maximal coherences. P takes value $0$ for the orthogonal situation, which gives no interference. Intermediate p values on the
$[-1, 1]$ segment are also possible. The extremes of maximal and minimal coherences deserve special attention.

Similarly, we can derive the dynamical equations in the format of density matrix. In the rotating-wave approximation, the semi-classical
time-dependent interaction Hamiltonian that describes the atom-laser coupling for this $\Lambda$ system is given by
\begin{equation}
H_{int}=-\hbar\left( \Omega_{1}e^{-i\Delta t}|c\ket\bra v_1|+\Omega_{2}e^{i\Delta t}|c\ket\bra v_2|+H.C.\right). \ee

The time evolution of the density matrix is given by the master equation
\begin{equation}
\dot{\rho}=-\frac{i}{\hbar}[H_{int},\rho]+L\rho, \label{drho}
\end{equation}
where $L\rho=L_1\rho+L_2\rho$ describes spontaneous emission terms. The spontaneous decay rate between two levels $|1\ket$,$|2\ket $is given by
\begin{equation}
\gamma=\frac{1}{4\pi\epsilon_0}\frac{4\omega^3\wp_{12}^2}{3 \hbar c^3}=\frac{\omega^3}{3\pi\hbar c^3 \epsilon_0}\wp_{12}^2.
\end{equation}
Let $\gamma'=\frac{\omega^3}{3\pi\hbar c^3 \epsilon_0}$, then $\gamma=\gamma'\wp_{12}^2$, thus the relaxation terms become
\begin{equation}
\begin{aligned}
L_1\rho=&-\gamma'[(\wp_{cv_1}\sigma_1^++\wp_{cv_2}\sigma_2^+)(\wp_{cv_1}\sigma_1+\wp_{cv_2}\sigma_2)\rho\\ &+\rho(\wp_{cv_1}\sigma_1^++\wp_{cv_2}\sigma_2^+)(\wp_{cv_1}\sigma_1+\wp_{cv_2}\sigma_2)\\
&-2(\wp_{cv_1}\sigma_1+\wp_{cv_2}\sigma_2)\rho(\wp_{cv_1}\sigma_1^++\wp_{cv_2}\sigma_2^+)];\label{rl11}
\end{aligned}
\end{equation}
\begin{equation}
L_2\rho=-\Gamma[\sigma_3^+\sigma_3\rho+\rho\sigma_3^+\sigma_3-2\sigma_3\rho\sigma_3^+]. \label{rl12}
\end{equation}

Here $\sigma_1^+=|v_1\ket \bra R_v|$, $\sigma_2^+=|v_2\ket \bra R_v|$, and $\sigma_3^+=|c\ket \bra R_c|$ are the atomic transition operators. We have
taken into consideration the interference introduced by the two decays from the ground state doublet to the same continuum.

Expanding Eq.(\ref{drho}) on the basis of $|c\ket$, $|v_1\ket$, $|v_2\ket$, $|R_c\ket$, $|R_v\ket$, and using the relaxation
Eqs.(\ref{rl11})-(\ref{rl12}), we obtain the dynamical evolution of the density matrix elements as,
\begin{align}
\dot{\rho_{11}}&=i\Omega_1^*\rho_{c1}-i\Omega_1\rho_{1c}-2\gamma_1\rho_{11}-p\sqrt{\gamma_1\gamma_2}(\rho_{12}+\rho_{21}),\label{P11} \\
\dot{\rho_{22}}&=i\Omega_2^*\rho_{c2}-i\Omega_2\rho_{2c}-2\gamma_2\rho_{22}-p\sqrt{\gamma_1\gamma_2}(\rho_{12}+\rho_{21}),\label{P22} \\
\dot{\rho_{Rc}}&=2\Gamma\rho_{cc},   \\
\dot{\rho_{Rv}}&=2\gamma_1\rho_{11}+2\gamma_2\rho_{22}+2p\sqrt{\gamma_1\gamma_2}(\rho_{12}+\rho_{21}),
\end{align}
and the non-diagonal terms
\begin{align}
\dot{\rho_{12}}=i\Omega_1^*\rho_{c2}-i\Omega_2\rho_{1c}-p\sqrt{\gamma_1\gamma_2}(\rho_{11}+\rho_{22})-\Gamma_{12}\rho_{12},\label{P12}\\
\dot{\rho_{1c}}=i\Omega_1^*(\rho_{cc}-\rho_{11})-i\Omega_2^*\rho_{12}-p\sqrt{\gamma_1\gamma_2}\rho_{2c}-\Gamma_{1c}\rho_{1c},\label{P1c}\\
\dot{\rho_{2c}}=i\Omega_2^*(\rho_{cc}-\rho_{22})-i\Omega_1^*\rho_{21}-p\sqrt{\gamma_1\gamma_2}\rho_{1c}-\Gamma_{2c}\rho_{2c}, \label{P2c}
\end{align}
where
\be
\begin{aligned}
\Gamma_{12}&=\gamma_1+\gamma_2+2i\Delta, \\ \Gamma_{1c}&=\Gamma+\gamma_1+i\Delta,\\ \Gamma_{2c}&=\Gamma+\gamma_2-i\Delta,
\end{aligned}
\ee are complex dephasing. Here we already exploit the norm preserving condition ($||\rho|| = 1$) of a density matrix for a closed system, so the
time derivation of $\rho_{cc}$ is not needed.

Defining the probability of emission as a sum of the population in the levels $v_1$, $v_2$ and $R_v$ for the system in Fig.\ref{f1}: \be
P_{\text{emiss}}(t)=1-\rho_{cc}(t)-\rho_{R_cR_c}(t),\label{pem} \ee with initial conditions $v_{1,2}(0)=0$, $c(0)=1$, and taking into account the
evolution of level $R_c$: $\dot{\rho}_{R_cR_c}(t)=2\Gamma{\rho}_{cc}(t)$, the probability  of emission is: \be
P_{\text{emiss}}(t)=1-\rho_{cc}(t)-2\Gamma\int_0^t\rho_{cc}(t')dt'\label{Peq}. \ee

Similarly, the probability of absorption is given by \be
P_{\text{abs}}(t)=\rho_{cc}(t)+\rho_{R_cR_c}(t)=\rho_{cc}(t)+2\Gamma\int_0^t\rho_{cc}(t')dt'.\label{Paq} \ee

Eqs.(\ref{P11})-(\ref{P2c}) and Eqs.(\ref{Peq})-(\ref{Paq}) are the formulas we will exploit to investigate the photovoltaics system. In the
following section III, $ode45$ function of MATLAB is used to solve these ordinary differential equations.

\section{Numerical simulations}

\begin{figure}[ht]
\begin{minipage}[b]{0.45\linewidth}
\centering
\includegraphics[width=\textwidth]{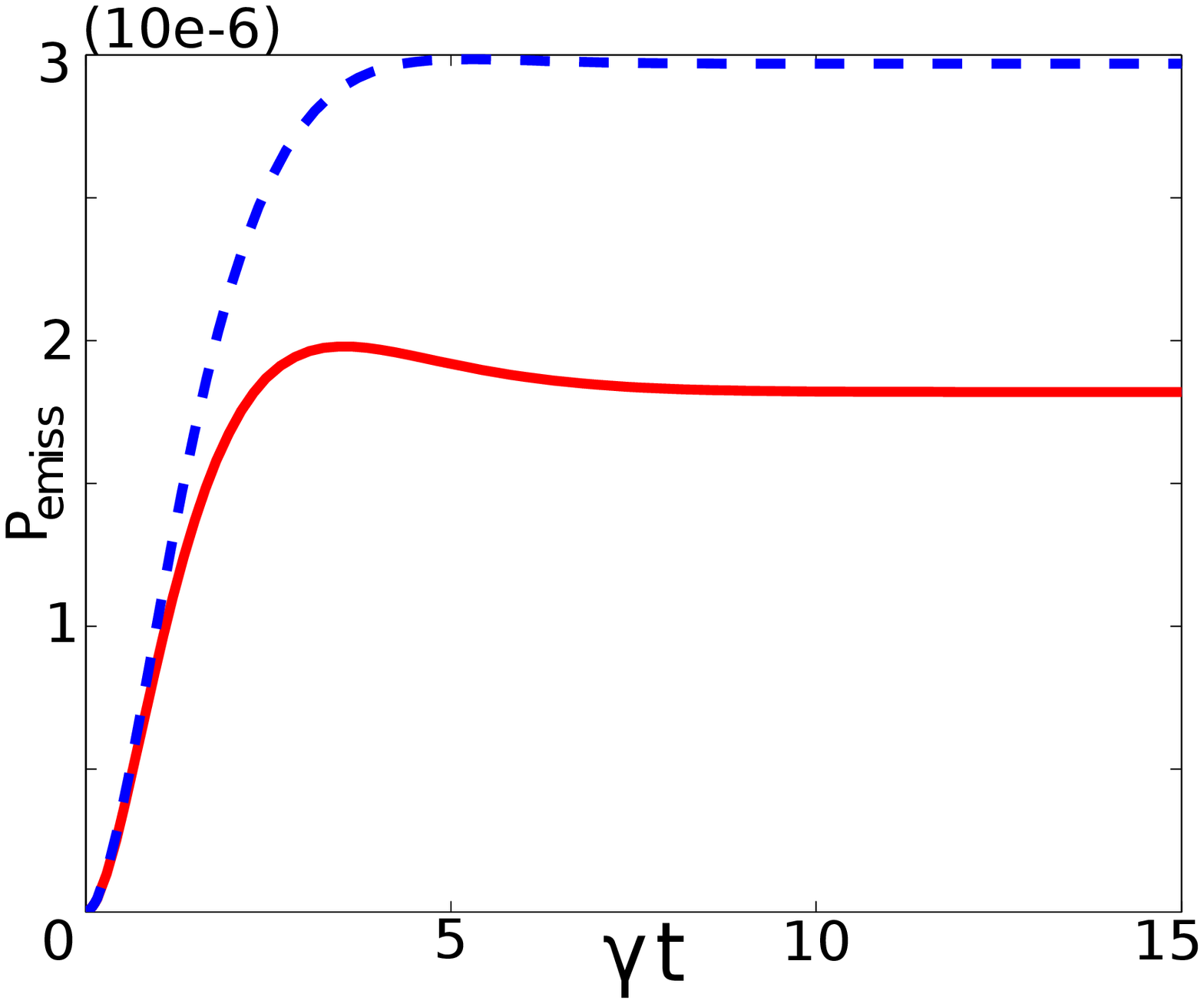}
\end{minipage}
\hspace{0.1cm}
\begin{minipage}[b]{0.45\linewidth}
\centering
\includegraphics[width=\textwidth]{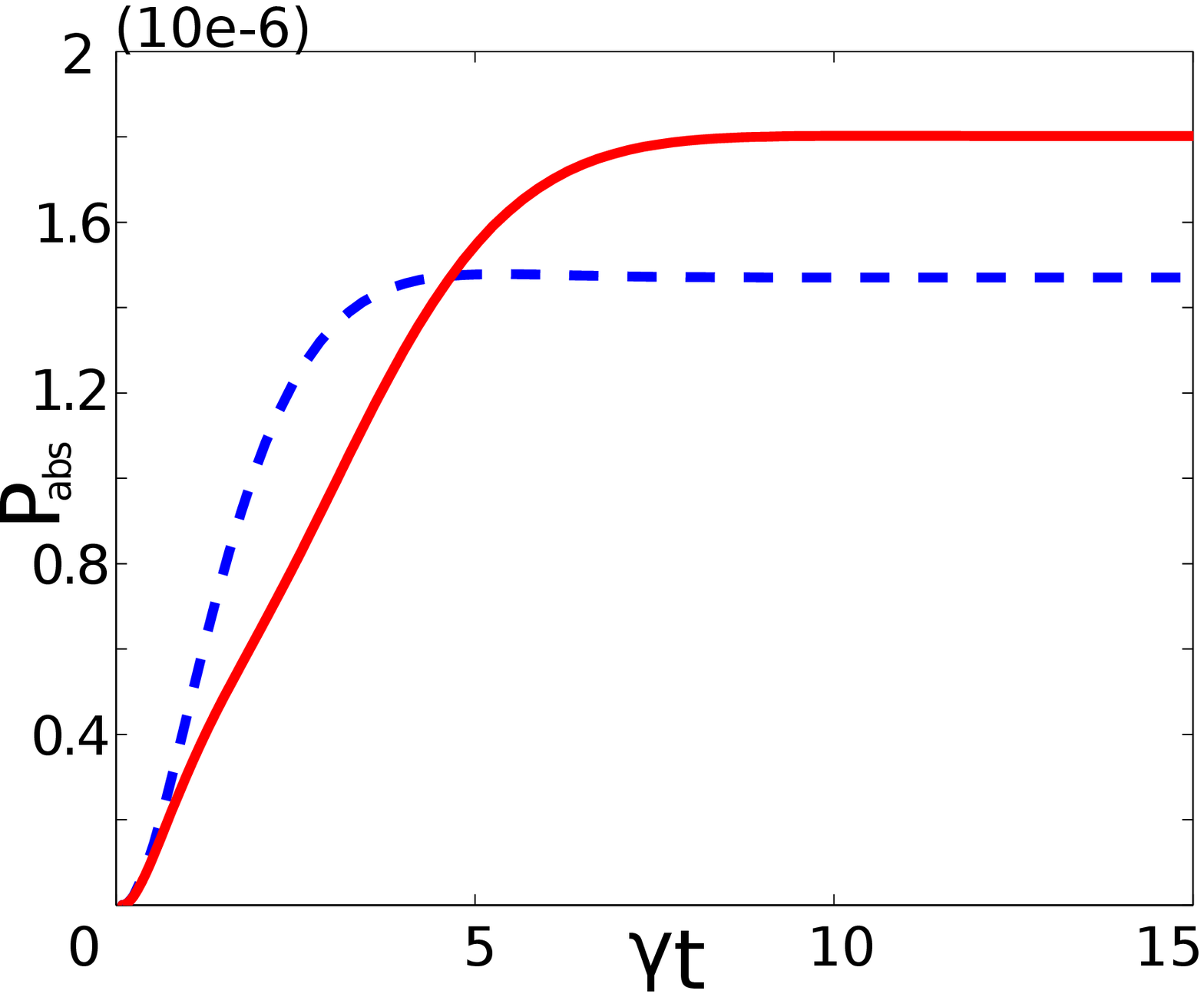}
\end{minipage}
\caption{Time dependence of probability of emission and absorption. $p=1$ (solid red), $p=0$ (dashed blue) with $\Omega_1=\Omega_2=0.001, \Delta=0.6,
\Gamma=1$, and $\gamma=1$.} \label{PP}
\end{figure}

Fig.\ref{PP} shows the dependence of probability of emission and absorption on time with and without interference. The red solid curves correspond to the maximal coherence with
$p=1$ and the blue dashed curves correspond to no coherence with $p=0$. The parameters we are using are Rabi frequency $\Omega_1=\Omega_2=0.001$,
detuning $\Delta=0.6$, and the spontaneous decay rates $\Gamma=\gamma=1$.

The effect of the Fano interference are apparent. We will take the steady states as an example. Without coherence, the probability of absorption is
$1.47\times 10^{-6}$, and the probability of emission is $2.94\times 10^{-6}$. With $p=1$, the probability of absorption is $1.8\times 10^{-6}$ and
the probability of emission is $1.8\times 10^{-6}$. Due to the fano interference, we obtain a nearly $22\%$ increase in absorption and a $38\%$
decrease in emission. The result meets our expectation that the balance between absorption and emission has been broken by the coherence induced by
the spontaneous decays. The output power of photovoltaic is proportional to the ratio of the population density on the excited state to that on the
ground state\cite{kk19}. Our result provides a possible method to enhance the power output of photovoltaic.

We should also expect that the detuning plays a role in changing the probability of emission and absorption. Fig.\ref{PPsDelta} shows the same plots
as shown in Fig.\ref{PP} but with a smaller detuning $\Delta=0.1$.  Under the same conditions, the probabilities of both emission and absorption
increase with this smaller detuning. As with larger detuning, the interference plays a similar role, giving an increase of $30\%$  in absorption and
a decrease of $34\%$ in emission.

\begin{figure}[ht]
\begin{minipage}[b]{0.45\linewidth}
\centering
\includegraphics[width=\textwidth]{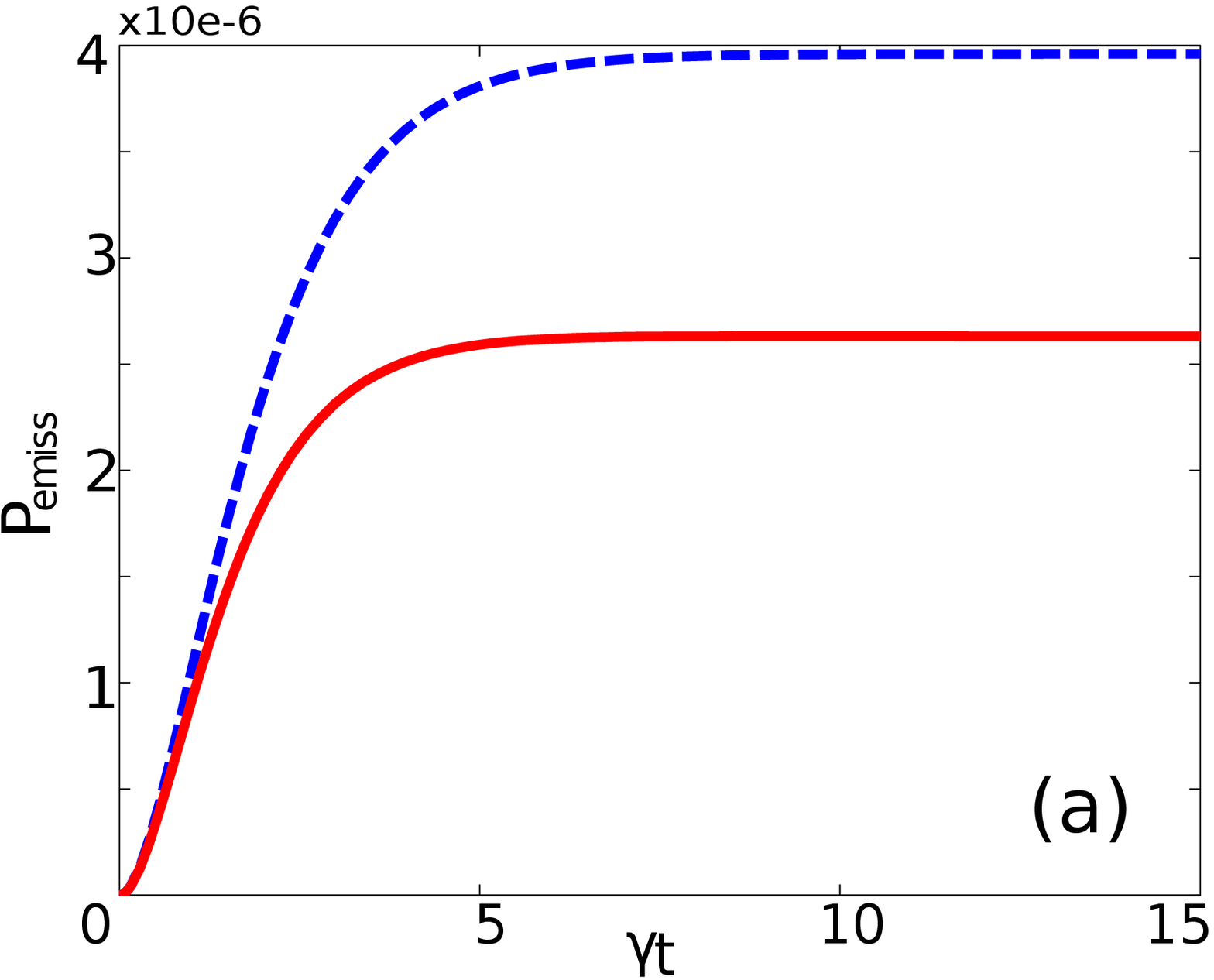}
\end{minipage}
\hspace{0.1cm}
\begin{minipage}[b]{0.45\linewidth}
\centering
\includegraphics[width=\textwidth]{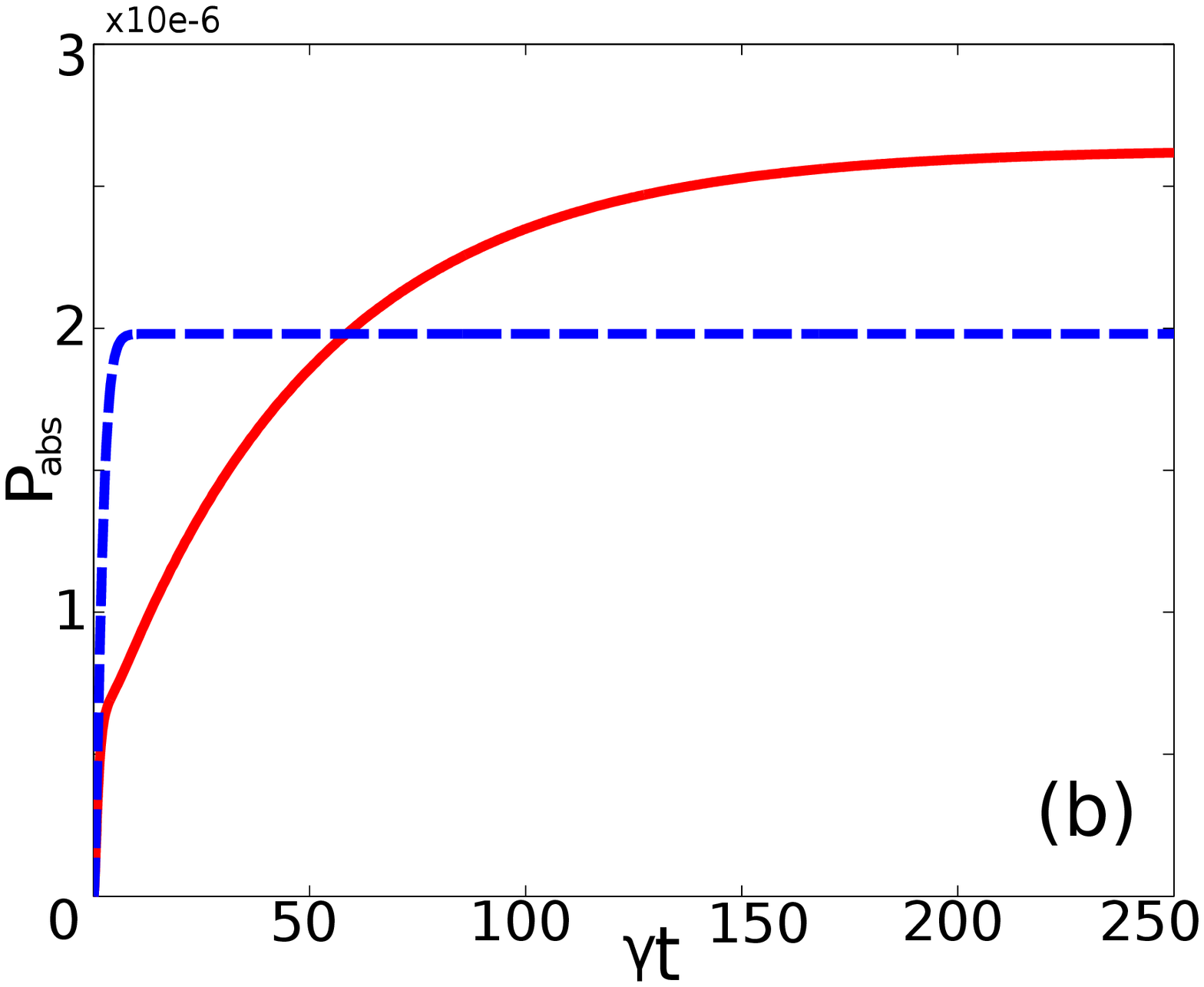}
\end{minipage}
\caption{Time dependence of probability of emission (a) and absorption (b). $p=1$ (solid red), $p=0$ (dashed blue) with $\Omega_1=\Omega_2=0.001,
\Delta=0.1, \Gamma=1$, and $\gamma=1$.} \label{PPsDelta}
\end{figure}

\section{Theoretical analysis}
To simplify the calculation, we assume a special case where $\gamma_1=\gamma_2=\gamma$. We will solve this problem analytically using both the
probability amplitude and population density approaches.

\subsection{Probability amplitude approach}
We borrowed the method Scully\cite{kk31} exploited and modified it to fit our model. The two extremes of maximal and minimal coherences have special
properties. We will only consider these two situations in our analysis. For the maximal coherence, we have $p=1$, the dynamical
Eqs.(\ref{v2})-(\ref{cc}) become
\begin{align}
\dot{v}_2&=-(\gamma+i\Delta)v_2-\gamma v_1-i\Omega_2c \label{sv2},\\ \dot{v}_1&=-(\gamma-i\Delta)v_1-\gamma v_2-i\Omega_1c, \label{sv1}\\
\dot{c}&=-i\Omega_2v_2-i\Omega_1v_1-\Gamma c. \label{scc}
\end{align}
Writing Eqs.(\ref{sv2})-(\ref{scc}) in matrix form, we obtain
\begin{equation}
\frac{d}{d\tau} \left(
  \begin{array}{c}
    v_2 \\
    v_1 \\
    c \\
  \end{array}
\right) =-V_f \left(
                   \begin{array}{c}
                     v_2 \\
                     v_1 \\
                     c \\
                   \end{array}
                 \right)-iV_p\left(
                             \begin{array}{c}
                               v_2 \\
                               v_1 \\
                               c \\
                             \end{array}
                           \right),
\end{equation}
where $\tau = \gamma t $, and the Fano decay matrix is defined by
\begin{equation}
V_f= \left(
  \begin{array}{ccc}
    1+i \tilde{\Delta} & 1 & 0 \\
    1 & 1-i \tilde{\Delta} & 0 \\
    0 & 0 & \tilde{\Gamma} \\
  \end{array}
\right),
\end{equation}
and the probe-field interaction is given by
\begin{equation}
V_p= \left(
  \begin{array}{ccc}
    0 & 0 & \tilde{\Omega}_2 \\
    0 & 0 & \tilde{\Omega}_1 \\
    \tilde{\Omega}_2 & \tilde{\Omega}_1 & 0 \\
  \end{array}
\right),
\end{equation}
with $\tilde{\Delta}=\frac{\Delta}{\gamma}$ and $\tilde{\Omega}_{1,2}= \frac{\Omega_{1,2}}{\gamma}$, $\tilde{\Gamma}=\frac{\Gamma}{\gamma}$.

It is intuitive to introduce a dressed basis in which the Fano coupling is transformed away. We proceed from the bare basis via the $U$, $U^{-1}$
matrices of diagonalization.
\begin{align}
U^{-1} &=\frac{1}{\sqrt{2}} \left(
                           \begin{array}{ccc}
                             1 & 1 & 0 \\
                             x-i \tilde{\Delta} & -x-i \tilde{\Delta} & 0 \\
                             0 & 0 & \sqrt{2} \\
                           \end{array}
                         \right), \\
U &=\frac{1}{\sqrt{2}x} \left(
                           \begin{array}{ccc}
                             x+ i \tilde{\Delta} & 1 & 0 \\
                             x-i \tilde{\Delta} & -1 & 0 \\
                             0 & 0 & \sqrt{2}x \\
                           \end{array}
                         \right),
\end{align}
where $x=\sqrt{1-\tilde{\Delta}^2}$.

The transformed state vector is defined by
\begin{equation}
U\left(
   \begin{array}{c}
     v_2 \\
     v_1 \\
     c \\
   \end{array}
 \right) =\left(
            \begin{array}{c}
              A_+ \\
              A_- \\
              B \\
            \end{array}
          \right),
\label{transform}
\end{equation}
which implies
\begin{equation}
\left(
            \begin{array}{c}
              \dot{A}_+ \\
              \dot{A}_- \\
              \dot{B} \\
            \end{array}
   \right)
   =-W_p \left(
            \begin{array}{c}
              A_+ \\
              A_- \\
              B \\
            \end{array}
          \right)-iW_f \left(
            \begin{array}{c}
              A_+ \\
              A_- \\
              B \\
            \end{array}
          \right), \label{dynamicaldress}
\end{equation}
where the diagonal operator is
\begin{equation}
W_f= U V_f U^{-1}=\left(
                          \begin{array}{ccc}
                            1+x & 0 & 0 \\
                            0 & 1-x & 0 \\
                            0 & 0 & \tilde{\Gamma} \\
                          \end{array}
                        \right),
\end{equation}
and the transformed interaction potential is
\begin{equation}
W_p= UV_p U^{-1}=\frac{1}{\sqrt{2}}\left(
  \begin{array}{ccc}
    0 & 0 & \tilde{\Omega}_+\\
    0 & 0 & \tilde{\Omega}_-\\
    \tilde{\Omega}'_+ & \tilde{\Omega}'_- & 0 \\
  \end{array}
\right),
\end{equation}
where
\begin{equation}
\begin{aligned}
\tilde{\Omega}_\pm&=[\tilde{\Omega}_2(x\pm i \tilde{\Delta})\pm\tilde{\Omega}_1]/x, \\
\tilde{\Omega}'_\pm&=[\tilde{\Omega}_2\pm\tilde{\Omega}_1 (x\mp i\tilde{\Delta})]/x.
\end{aligned}
\end{equation}
The equations of motion in terms of $A_\pm$ and $B$ are then
\begin{align}
\frac{d A_+}{d\tau}&=-(1+x)A_+-\frac{i}{\sqrt{2}}\tilde{\Omega}_+B, \label{dAp} \\
\frac{d A_-}{d\tau}&=-(1-x)A_--\frac{i}{\sqrt{2}}\tilde{\Omega}_-B, \label{dAn} \\
\frac{d B}{d\tau}&=-\frac{i}{\sqrt{2}}[\tilde{\Omega}'_+A_+ +\tilde{\Omega}'_-A_-]-\tilde{\Gamma}B. \label{dB}
\end{align}

Note that the transformed interaction matrix in Eq. (\ref{dynamicaldress}) is not symmetric, i.e., it is non-Hermitian, thus the absorption-emission
balance is broken.

From Eqs.(\ref{dAp})-(\ref{dB}), we can derive the analytical solutions of probability amplitude in the dressed states. To find the probability
amplitude for emission, we take the initial condition as $B(0)=1$ and $A_\pm(0)=0$, and we assume the Rabi frequency of the driving fields $\W_{1,2}$
to be weak enough to apply the perturbation method. The first order approximation of $B$ is a pure exponential function $B^{(0)}\cong
e^{-\tilde{\Gamma}\tau}$. According to our numerical simulations, this equation only works well for a short time in the beginning. A higher order
approximation is needed. Replacing $B$ in both Eqs.(\ref{dAp})-(\ref{dAn}) with $B^{(0)}$, we obtain
\begin{equation}
\begin{aligned}
\frac{d A_+^{(1)}}{d\tau}=&-(1+x)A_+^{(1)}-\frac{i}{\sqrt{2}}\tilde{\Omega}_+e^{-\tilde{\Gamma}\tau},\\
\frac{d A_-^{(1)}}{d\tau}=&-(1-x)A_-^{(1)}-\frac{i}{\sqrt{2}}\tilde{\Omega}_-e^{-\tilde{\Gamma}\tau},
\end{aligned}
\end{equation}
giving the $A_\pm$
\begin{equation}
\begin{aligned}
A_\pm^{(1)}\cong& -\frac{i}{\sqrt{2}}\tilde{\Omega}_\pm\int_0^t e^{-(1\pm x-\tilde{\Gamma}\tau)(t-t')}dt'\\
=&-i\frac{e^{-\tilde{\Gamma}\tau}-e^{-(1\pm x)t}}{\sqrt{2}(1\pm x-\tilde{\Gamma}\tau)}\tilde{\Omega}_+. \label{Apn}
\end{aligned}
\end{equation}

Substituting Eq.(\ref{Apn}) into Eq.(\ref{dB}), we obtain the next order of $B$, \be B^{(1)}=(a_0\tau-a_+-a_-+1) e^{-\tilde{\Gamma}
\tau}+a_+e^{-(1+x) \tau}+a_-e^{-(1-x) \tau}, \ee
where \be \begin{aligned}
a_0 &=\frac{x^2\tilde{\Omega}'_+\tilde{\Omega}_+}{1+x-\tilde{\Gamma}}+\frac{x^2\tilde{\Omega}'_-\tilde{\Omega}_-}{1-x-\tilde{\Gamma}}\\
a_\pm &=\tilde{\Omega}'_\pm\tilde{\Omega}_\pm\frac{x^2}{(1+x-\tilde{\Gamma})^2}.
\end{aligned} \ee

Similarly, we can get the probability amplitude for absorption. There is no population in the ground state, and the initial condition is $c(0)=0$,
$v_1=1$ or $v_2=1$.  Eq.(\ref{transform}) gives $B(0)= 0$, and \be A_\pm =\frac{1}{\sqrt{2}x}[v_2 (x\pm i\tilde{\Delta})\pm v_1]. \ee

We need to consider $v_1=1$ and $v_2=1$ separately.

For $v_1=1$, $A_{\pm} =\pm\frac{1}{\sqrt{2}x}$. In the situation of a weak field, we have the first order approximation of $A_\pm$, \be
A_\pm^{(0)}=\pm\frac{1}{\sqrt{2}x}e^{-(1\pm x)}. \label{Apn1}\ee

Substituting Eq.(\ref{Apn1}) into Eq.(\ref{dB}), the next order of $B$ is \be
B^{(1)}=(b_+-b_-)e^{-\tilde{\Gamma}\tau}-b_+e^{-(1+x)\tau}+b_-e^{-(1-x)\tau}, \ee
where
\begin{equation}
b_\pm=-i\frac{\tilde{\Omega}'_\pm}{2(1\pm x-\tilde{\Gamma})}.
\end{equation}

For $v_2=1$, we have $A_{\pm} =\pm\frac{x\pm i\tilde{\Delta}}{\sqrt{2}x}$. Compared with the situation $v_1=1$, there is only a time-independent
coefficient difference. We can directly obtain
\begin{equation}
\begin{aligned}
B^{(1)}&=(b_+(x+i\tilde{\Delta})-b_-(x-i\tilde{\Delta}))e^{-\tilde{\Gamma}\tau} \\
&+b_+e^{-(1+x)\tau}(x+i\tilde{\Delta})+b_-e^{-(1-x)\tau}(x-i\tilde{\Delta}),
\end{aligned}
\end{equation}
where $b_\pm$ are the same as defined for $v_1=1$.

For the situation without coherence, or $p=0$, the calculation is straightforward. No dressed states are needed. Eqs.(\ref{v2})-(\ref{cc}) become
\begin{align}
\dot{v}_2&=-(\gamma+i\Delta)v_2-i\Omega_2c \label{ssv2}, \\
\dot{v}_1&=-(\gamma-i\Delta)v_1-i\Omega_1c \label{ssv1}, \\
\dot{c}&=-i\Omega_2v_2-i\Omega_1v_1-\Gamma c \label{sscc}.
\end{align}

For emission, we have the same initial condition of $c(0)=1$, $v_{1,2}(0)=0$.  The first order approximations are \be
\begin{aligned}
c^{(0)}&=e^{-\tilde{\Gamma} t},\\
v_1^{(0)}&=\frac{i\tilde{\Omega_1}}{\tilde{\Gamma}-\tilde{\Gamma_1}}(e^{-\tilde{\Gamma} t}-e^{-\tilde{\Gamma_1} t}),\\
v_2^{(0)}&=\frac{i\tilde{\Omega_2}}{\tilde{\Gamma}-\tilde{\Gamma_2}}(e^{-\tilde{\Gamma} t}-e^{-\tilde{\Gamma_2} t}),
\end{aligned} \label{v1v2}\ee
where
\begin{equation}
\begin{aligned}
\tilde{\Gamma_1}=1-i\tilde{\Delta},\\
\tilde{\Gamma_2}=1+i\tilde{\Delta}.
\end{aligned}
\end{equation}

Substituting Eq.(\ref{v1v2}) into Eq.(\ref{sscc}), we get the next order of probability amplitude,
\begin{equation}
c^{(1)}=(a_0t-a_1-a_2+1) e^{-\tilde{\Gamma} t}+a_1e^{-\tilde{\Gamma_1} t}+a_2e^{-\tilde{\Gamma_2} t},
\end{equation}
where
\begin{equation}
\begin{aligned}
a_0 &=(\frac{\tilde{\Omega_1}^2}{\tilde{\Gamma}-\tilde{\Gamma_1}}+\frac{\tilde{\Omega_2}^2}{\tilde{\Gamma}-\tilde{\Gamma_2}}), \\
a_1 &=\frac{\tilde{\Omega_1}^2}{(\tilde{\Gamma}-\tilde{\Gamma_1})^2}, \\
a_2 &=\frac{\tilde{\Omega_2}^2}{(\tilde{\Gamma}-\tilde{\Gamma_2})^2}.
\end{aligned}
\end{equation}

At large time $\tau\gg1,1/\tilde{\Gamma}$ and with weak field approximation, by applying Eq.(\ref{Peq}), we can obtain the probability of emission
\begin{equation}
P_{\text{emiss}}\simeq
\frac{(4\Delta^2(3\Gamma-1)+(\Gamma^2-1)(3\Gamma+1))(4\Omega_1^2+4\Omega_2^2)}{(4\Delta^2+1)^2\Gamma+2(4\Delta^2-1)\Gamma^3+\Gamma^5}.
\end{equation}
For absorption, we only show the initial condition $v_1(0)=1$, $c(0)=0$ as an example. Following the same process for emission, we obtain
\begin{equation}
\begin{aligned}
v_1^{(0)}&=e^{-\Gamma_1 t},\\
c^{(0)}&=\frac{i\Omega_1}{\Gamma_1-\Gamma}\left(e^{-\Gamma_1 t}-e^{-\Gamma t}\right),\\
v_2^{(0)}&=\frac{\Omega_1\Omega_2^*}{\Gamma_1-\Gamma}\left(\frac{e^{-\Gamma t}-e^{-\Gamma_2 t}}{\Gamma_2-\Gamma}+\frac{e^{-\Gamma_1 t}-e^{-\Gamma_2
t}}{\Gamma_1-\Gamma_2}\right).
\end{aligned}
\end{equation}
Therefore
\be
\begin{aligned}
c^{(1)}=&-ib_0e^{-\Gamma t}-\frac{ib_1}{\Gamma-\Gamma_1}(e^{-\Gamma_1 t}-e^{-\Gamma t}) \\
&-\frac{ib_2}{\Gamma-\Gamma_2}(e^{-\Gamma_2 t}-e^{-\Gamma t}),
\end{aligned}
\end{equation}
where
\begin{equation}
\begin{aligned}
b_0&=\Omega_1|\Omega_2|^2\frac{1}{(\Gamma_1-\Gamma)(\Gamma_2-\Gamma)},\\
b_1&=\Omega_1|\Omega_2|^2\left(\frac{1}{(\Gamma_1-\Gamma)(\Gamma_1-\Gamma_2)}+\frac{1}{|\Omega_2|^2}\right),\\
b_2&=\Omega_1|\Omega_2|^2\frac{1}{(\Gamma_2-\Gamma)(\Gamma_1-\Gamma_2)}.
\end{aligned}
\end{equation}
By applying Eq.(\ref{Paq}), the probability of absorption is given by \be
\begin{aligned}
P_{\text{abs}}(\tau|c)&=\rho_{cc}(\tau)+2\tilde{\Gamma}\int_0^\tau\rho_{cc}(\tau')d\tau'\\
&\simeq \frac{4\Omega_1^2(1+\Gamma)}{4\Delta^2+(\Gamma+1)^2}.
\end{aligned} \ee

\subsection{Density matrix approach}

It will be interesting to derive the analytical solution of the probability of emission and absorption in the density element form. However, we found
this too complex with the existence of coherence. Therefore, we take the special situation of no interference($p=0$) as an example. The density
matrix becomes
\begin{align}
\dot{\rho_{11}}&=i\Omega_1^*\rho_{c1}-i\Omega_1\rho_{1c}-\gamma\rho_{11}, \label{P011}\\
\dot{\rho_{22}}&=i\Omega_2^*\rho_{c2}-i\Omega_2\rho_{2c}-\gamma\rho_{22}, \label{P022}\\
\dot{\rho_{cc}}+\dot{\rho_{Rc}}&=-i\Omega_1(\rho_{c1}-\rho_{1c})-i\Omega_2(\rho_{c2}-\rho_{2c}),\label{P0cd}\\
\dot{\rho_{12}}&=i\Omega_1^*\rho_{c2}-i\Omega_2\rho_{1c}-\Gamma_{12}\rho_{12}, \label{P012}\\
\dot{\rho_{1c}}&=i\Omega_1^*(\rho_{cc}-\rho_{11})-i\Omega_2^*\rho_{12}-\Gamma_{1c}\rho_{1c}, \label{P01c}\\
\dot{\rho_{2c}}&=i\Omega_2^*(\rho_{cc}-\rho_{22})-i\Omega_1^*\rho_{21}-\Gamma_{2c}\rho_{2c}. \label{P02c}
\end{align}
Notice that Eq.(\ref{P0cd}) is already the time derivation of the core of the probability of emission and absorption formulas. It is the key to solve
these equations. With the initial condition of $\rho_{cc}=1$, Eq.(\ref{P01c}),Eq.(\ref{P02c}) will give us the same time evolution of $\rho_{1c}$ and
$\rho_{2c}$, thus
\begin{equation}
\rho_{1c}=\frac{i\Omega_1}{\Gamma_{13}-\gamma}(e^{-\gamma t}-e^{-\Gamma_{13}t}).
\end{equation}
Integrating on both sides of Eq.(\ref{P0cd}) from zero to infinity, and substituting the results into Eq.(\ref{pem}), we get the probability of
emission
\begin{equation}
P_{emiss}=\frac{4(\Omega_1^2+\Omega_2^2)}{4\Delta^2+\Gamma^2} \frac{(4\Delta^2+\Gamma^2)\Gamma+\Gamma^2-4\Delta^2}{4\Delta^2+(\Gamma+1)^2}.
\end{equation}

For absorption,  we have $\rho_{11}(0)=1$; substituting it into Eq.(\ref{P011}), we get $\rho_{11}(t)=e^{-\gamma t}$. From Eq. (\ref{P01c}), we get
\begin{equation}
\rho_{1c}=\frac{i\Omega_1}{\gamma-\Gamma_{13}}(e^{-\gamma t}-e^{-\Gamma_{13}t}).
\end{equation}
According to our numerical simulations, $\rho_{2c}$ is always much smaller than $\rho_{1c}$, we can ignore the contribution by $\rho_{2c}$. Thus
\begin{equation}
\begin{aligned}
\dot{\rho_{cc}}+\dot{\rho_{Rc}}&=-i\Omega_1(\rho_{c1}-\rho_{1c})\\
&=\frac{\Omega_1^2}{\Delta^2+\Gamma^2}\{2\Gamma e^{-\gamma t} \\
&-e^{-(\Gamma+\gamma)t}(2\Gamma cos(\Delta t)+2\Delta sin(\Delta t)\}.
\end{aligned}\label{ing}
\end{equation}
Again integrating on both sides of Eq.(\ref{ing}) from zero to infinity, and substituting the result into Eq.(\ref{Paq}), we get
\begin{equation}
P_{abs}=\frac{4\Omega_1^2}{4\Delta^2+\Gamma^2} \frac{(4\Delta^2+\Gamma^2)\Gamma+\Gamma^2-4\Delta^2}{4\Delta^2+(\Gamma+1)^2}.
\end{equation}

To verify the analytical solutions, we can compare them with the numerical results. Let's take the same parameters we have used in Fig.\ref{PP},
except that here we have $p=0$. In the steady states, we have $P_{emiss}=3.992\times 10^{-6}$ for analytical solution and $P_{emiss}=3.998\times
10^{-6}$ for numerical solution, and $P_{abs}=1.996\times 10^{-6}$ for analytical solution and $P_{abs}=1.999\times 10^{-6}$ for numerical solution.
Both sets match well, showing that these two analytical solutions are good enough to describe the emission and absorption dependence on the system
parameters.

\section{Conclusion}
We have studied the effect of the Fano interference in a three-level system with reservoir. We found that the balance between emission and absorption
for the original system has been broken, because the interference largely suppresses the emission process and enhances the absorption process. This
property can possibly be applied to improve the efficiency of solar cell, as the interference can increase the probability of absorption and decrease
the probability of emission. Under the weak field approximation, the analytical solutions of probability amplitude and density elements are derived.
The results matched well with our numerical simulations.

\section{Acknowledgement}
Authors thank Marlan Scully for his support on this project, and thanks Sumanta Das, Konstantin E. Dorfman and Pankaj Jha for helpful discussions.
This project is supported by the Robert A. Welch Foundation (Grant No. A-1261), D. S. thanks the support from the Fujian Natural Science
Foundation(2018I0019), Science Technology innovation of Xiamen(3502Z20183062),Y.R. gratefully acknowledge the support from The Advanced Materials and
Manufacturing Processes Institute at the University of North Texas seed research project and from the UNT Research Initiation Grant..


\begin{thebibliography}{10}

\bibitem{CPT} E. Arimondo, Progress in Optics, edited by E. Wolf(Elsevier Science, Amsterdam, 1996), Vol. XXXV, p. 257.

\bibitem{book} M. O. Scully and M. S. Zubairy, Quantum Optics (Cambridge University Press, Cambridge, England, 1997).

\bibitem{harris97phys2day} S. E. Harris, Phys. Today {\bf 50}, 36 (1997).

\bibitem{harris93prl} S. E. Harris, Phys. Rev. Lett. {\bf 70}, 552 (1993).

\bibitem{fleisch05rmp} M. Fleischhauer, A. Imamoglu, and J. P. Marangos, Rev. Mod. Phys. {\bf 77}, 633 (2005).

\bibitem{ok86jetp} O. Kocharovskaya, and Y. I. Khanin, Sov. Phys. JETP {\bf 63}, 945 (1986).

\bibitem{sau05pra} V.A. Sautenkov, Y.V. Rostovtsev, C. Y. Ye, G.R. Welch, O. Kocharovskaya, and M.O. Scully, Phys. Rev. A {\bf 71}, 063804 (2005).

\bibitem{scully91prl} M. O. Scully, Phys. Rev. Lett. {\bf 67}, 1855 (1991); M. O. Scully and M. Fleischhauer, \textit{ibid}, {\bf 69}, 1360  (1992);  A. S. Zibrov \textit{et. al.
ibid} {\bf 76}, 3935 (1996).

\bibitem{ye03optlett} C. Y. Ye, V. A. Sautenkov, Y. V. Rostovtsev, and M. O. Scully, Opt. Lett. {\bf 28}, 2213 (2003).

\bibitem{harris98prl} S. E. Harris, and Y. Yamamoto, Phys. Rev. Lett. {\bf 81}, 3611 (1998).

\bibitem{lukin00amop} M. D. Lukin, P. R. Hemmer, and M. O. Scully, Adv. in At. Mol. Opt. Phys. {\bf 42}, 347 (2000).

\bibitem{kk1} M.O. Scully and M. S. Zubairy Quantum Optics, (Cambridge Press, London 1997)

\bibitem{kk2} U. Fano, Phys. Rev. {\bf 124}, 1866 (1961).

\bibitem{kk3} G. S. Agarwal , Springer Tracts in Modern Physics: Quantum Optics (Springer-Verlag, Berlin, 1974).

\bibitem{kk5} S. E. Harris, Phys. Rev. Lett. {\bf 62}, 1033 (1989); M. O. Scully, S. -Y. Zhu, and A. Gavrielides, Phys. Rev. Lett. {\bf 62}, 2813 (1989).

\bibitem{kk6} O. Kocharovskaya, Phys. Rep. {\bf 219}, 175 (1992).

\bibitem{kk12} M. O. Scully, M. S. Zubairy, G. S. Agarwal, and H. Walther, Science {\bf 299}, 862 (2003).

\bibitem{kk8} A. Ishizaki, and G . Fleming, PNAS {\bf 106}, 17255 (2009).

\bibitem{kk13} P. W¡§urfel, Physics of Solar Cells, (Wiley-VCH Verlag gmbH \&Co., Weinheim, 2009).

\bibitem{kk10} M. O. Scully, S.Y. Zhu, and A. Gavrielides, Phys. Rev. Lett. {\bf 62}, 2813 (1989).

\bibitem{kk11} S.E. Harris, Phys. Rev. Lett. {\bf 62}, 1033 (1989).

\bibitem{kk14} M. O. Scully, Phys. Rev. Lett. {\bf 104}, 207701 (2010).

\bibitem{kk16} M. O. Scully Coherent Control, Fano Interference, and Non-Hermitian Interactions, Workshop held in May, 1999 (Kluwer Academic Publishers, Norwell, MA,
2001).

\bibitem{kk19} A. A. Svidzinsky, K. E. Dorfman, and M. O. Scully, Phys. Rev. A, {\bf 84}, 053818 (2011)

\bibitem{kk20} K. E. Dorfman, P. K. Jha, and Sumanta Das, Phy. Rev. A, {\bf 84}, 053803 (2011)

\bibitem{kk31} M. O. Scully, Advances in Multi-photon Processes and Spectroscopy, {\bf 14}, 126-132 (1999)

\end{thebibliography}
\end{document}